\documentclass[a4paper,eqsecnum,aps,epsf,twocolumn,10pt]{article}
\usepackage{graphicx}
\usepackage[usenames]{color}
\usepackage{bm}
\usepackage{theorem}

\begin{document}

\twocolumn[%
\begin{flushleft}
{\Large\bf
 A new set of variables in the three-body problem
} 
\end{flushleft}
\par\vspace{0.5cm} KENJI HIRO KUWABARA $^{1,2}$ and  KIYOTAKA TANIKAWA $^{2}$ 
\par\vspace{0.5cm} $^{1}$ Department of Applied Physics, Faculty of 
Science and Engineering\\ Waseda University, Tokyo 169-8555, Japan \\ 
\vspace{0.2cm} $^{2}$ National Astronomical Observatory of Japan, 
Mitaka, Tokyo 181-8588, Japan.

\date{\today}
\begin{center}
\begin{minipage}{0.8\linewidth}
%\begin{flushleft}
We propose a set of variables of the general three-body problem both for 
two-dimensional and three-dimensional cases. Variables are 
$(\lambda,\theta,\Lambda, \Theta,k,\omega)$ or equivalently 
$(\lambda,\theta,L,\dot{I},k,\omega)$ 
for the two-dimensional problem, and 
$(\lambda,\theta,L,\dot{I},k,\omega,\phi,\psi)$ 
for the three-dimensional problem. Here $(\lambda,\theta)$ and  
$(\Lambda,\Theta)$ specifies the positions  
in the shape spheres in the configuration and momentum spaces, 
$k$ is the virial ratio, $L$ is the total angular momentum, $\dot{I}$ is 
the time derivative of the moment of inertia, and  
$\omega,\phi$, and $\psi$ are the Euler angles to bring the momentum 
triangle from the nominal position to a given position.   
This set of variables defines a {\it shape space } 
of the three-body problem. 
This is also used as an initial condition space. 
The initial condition of the so-called free-fall three-body problem is 
$(\lambda,\theta,k=0,L=0,\dot{I}=0,\omega=0)$. 
We show that the hyper-surface $\dot{I} = 0$ is a global surface of section. 
%\end{flushleft}
\end{minipage}
\end{center}

\vspace{\baselineskip}

%\keywords{Three-body Problem, Angular momentum, 
%Time derivative of the inertial moment }

\vspace{0.3cm}

] 

\vspace{0.3cm}

%\keywords{Three-body Problem, Angular momentum, 
%Time derivative of the inertial moment }

\section{Introduction and motivation}

The three-body problem has been studied extensively 
by numerical simulations and by analytical methods.
However, in full generality this problem is of too high dimensionality 
and is too complicated for a systematic analysis. 
This has led so far to the study of various 
simplified and restricted versions of the problem. 
The free-fall problem (\cite{Agekyan}, \cite{Tanikawa1995},   
referred to as the FFP; the definition will be given later) 
is one of these. The isosceles problem  
(\cite{V.M.Alekseev}, \cite{Brou}, \cite{ZaCh}) and 
the rectilinear problem (\cite{HM1993}, \cite{TM2000a,TM2000b}) 
are other examples.  
A research looking at the whole phase space of the planar or
three-dimensional three-body problem is desirable.   

In the present work, we propose a new setting of the problem 
which is suitable for large-scale numerical studies, and which is
hopefully suitable for theoretical consideration. Our setting is 
summarized into a word 'shape space'. The shape space is a direct 
product of the shape sphere in the configuration space, 
the shape sphere in the momentum space, and their relative size and
orientation. This is not the phase space, 
but can be expanded to recover the phase space.  
%We are motivated from the expectation that, in near future, 
%we may be possible to explore the whole phase space, at least for 
%the planar case, of the three-body problem with the aid of fast computers.  

In getting the shape space, we have two guiding principles: 
equivalence relation and boundedness. 
We do not want to integrate orbits which can be transformed into one 
another by a suitable change of variables. 
These orbits are considered equivalent. 
We will consider only orbits of different equivalence classes.
Orbits belonging to different equivalence classes will be called 
independent. 
Boundedness is related to the size of the initial condition space.  
In order that a numerical study of the totality of the solution of the 
three-body problem be feasible, 
the initial condition space must be either bounded or finite.
Otherwise, it is impossible to exhaust the initial conditions  
using even the fastest computer.
So we would like to impose boundedness on the ranges of variables.

We consider the planar three-body problem since the extension to the three
dimensions is not so difficult. The phase space is $R^{12}$. 
As is well-known (\cite{Whittaker1952}, p.351), the possible minimum order of
differential equations is four. In order to attain this order, 
several processes are consecutively done. Thus, equations of motion are 
reduced from the 12th to the 8th order by using the four integrals of 
motion of the center of gravity. 
The use of the angular momentums reduces the order 
to 7, and  use of the elimination of the nodes reduces the order to 6.
Lastly, it is possible again to reduce the order of the equations 
to four by using the integral of energy and eliminating the time.

This conventional reduction, however, does not lead to a best choice of 
variables for the initial value problem. 
The space of the states of motion described by these 4th-order 
differential equations may have extremely complicated structure. 
Even if we stop the reduction at the 6th-order, variables 
have infinite ranges.

%So we here take a different approach. 
%For the computer simulation, one of the most important factors is 
%to minimize the number of calculations.
%In the case of the three-body problem, to solve one initial 
%condition does not need so much time if we use a good 
%software (\cite{MS1990}), and a hardware 
%as long as the orbit does not make a close approach to triple collision.
%So what we must reduce is the number of variables of the 
%initial condition space. Another important factor is the boundedness of
%variables. This enables us to keep the number of grid points (for
%numerical integrations) of the phase space finite.
%In addition, we want to keep clear the physical meaning of variables.  
%We put the FFP as a starting point. 
%The FFP is a very good problem in the sense of geometrical clarity.  
%We want to extend the FFP so that the FFP is included as a subspace. 

The concept of the shape sphere in the configuration space seems 
due to McGehee in the rectilinear three-body problem when 
he devised the variables now called after his name \cite{McGehee74}. 
Later Moeckel \cite{Moeckel88} explicitly introdcued the shape sphere 
in the planar three-body problem. Then, this notion is essentially used 
in \cite{eight2000} in order to obtain the figure-eight periodic
solution with variational techniques. Our shape space extends the notion 
to the whole phase space. 

We start in \S 2 by introducing the FFP and the involved study for it. 
In \S 3 we extend the FFP to any given mass ratio and the whole phase 
space in the planar case.
In \S 4 we suggest the existence of the 'semi' global surface of
section and generalize this results to the three dimensions.
In the final section, we summarize and discuss our results.

\section{The Free-Fall Problem}

\subsection{Definition}

The free-fall problem (FFP) is characterized by the  zero initial 
velocities, and has been extensively studied 
by Russian and Japanese schools(\cite{Agekyan},\cite{Anosova1986},
\cite{Tanikawa1995}, \cite{Umehara00}).
In the FFP, the total energy of the three bodies
$m_i,i=1,2,3$ is negative and their angular momentum is zero. 
We here consider the equal mass case: $m_1=m_2=m_3$. 
In this problem, motions starting from similar triangles transform into 
one another under appropriate changes of coordinates and time, 
so we identify these motions. Dissimilar triangles 
correspond to independent motions.

Let mass points $m_2$ and $m_3$ stand still at $A(-0.5,0.0)$ and 
$B(+0.5,0.0)$, respectively in the $(x,y)$ plane and $m_1$ stand still 
at a point $P(x,y)$ where
\begin{eqnarray}
 (x,y) \in {\cal D} =   \hspace{5cm} \\ \nonumber
 \{(x,y):x \ge 0,y \ge 0,(x+0.5)^2+y^2 \le 1 \} .
\end{eqnarray}
If $m_1$ changes position in ${\cal D}$, then triangles satisfying 
the condition  
$\overline{m_2m_3} \ge \overline{m_1m_2} \ge \overline{m_1m_3}$ are exhausted. 
Conversely, any triangle is similar to one of the triangles 
formed by three mass points $m_1$, $m_2$, and $m_3$ as above. 
Thus the positions of $P \in {\cal D}$ specify all
possible initial conditions. 

\subsection{An attempt to include velocities}

Anosova et al. (\cite{Anosova1981}) considered the region ${\cal D}$ 
of the free-fall problem. 
Then they supposed that the system rotates 
in the plane of initial triangle (2D problem) counterclockwise;
the velocity vectors of components A (distant component) and C 
(component inside ${\cal D}$) are orthogonal to their radius-vectors 
in the center-of-gravity coordinate system;
the angular momenta of these bodies are the same; 
the velocity of component B is given so that
the center-of-gravity of the triple system is motionless; 
the speed of rotation is parameterized by initial virial ratio $k$.
Thus the initial conditions are defined by three parameters:
coordinates $(x,y)$ of C component in the region ${\cal D}$ 
and virial ratio $k$. 

However, their formulation lost the boundedness of 
the initial configuration space.
This boundedness is one of the most important properties of the FFP.
What we should do is to recover this.

\section{ The definition of our variables for the planer case }

\subsection{Equations of motions for the planar three-body problem}

 Let $m_k >0$ be the masses of point particles with positions 
${\bf q}_k \in {\bf R}^2$ and momenta ${\bf p}_k \in {\bf R}^2;
k=1,2,3$.
 Let ${\bf q,p} \in {\bf R}^6$ denote the vectors 
$({\bf q}_1, {\bf q}_2, {\bf q}_3)$, $({\bf p}_1, {\bf p}_2, 
{\bf p}_3)$.
 The three-body problem is governed by the Hamiltonian function

\begin{eqnarray}
 H({\bf p},{\bf q}) =& 
\frac{1}{2}{\bf p} \cdot A^{-1} {\bf p} - U({\bf q})  = 
T ({\bf p}) - U({\bf q})  \label{eq:energy} \\ \nonumber 
\end{eqnarray}
where $A$ is the $6 \times 6$ mass matrix 
$diag(m_1, m_1, m_2, m_2, m_3, m_3)$, a dot denotes the scalar product in 
${\bf R}^6$, and 
\begin{eqnarray}
 U({\bf q})=\frac{m_1m_2}{|{\bf q}_1-{\bf q}_2|}+
\frac{m_2m_3}{|{\bf q}_2-{\bf q}_3|}+
\frac{m_3m_1}{|{\bf q}_3-{\bf q}_1|}.   \label{eq:potential}
\end{eqnarray}
Hamilton's equations are 
\begin{eqnarray}
 \dot{ {\bf q} }= A^{-1} {\bf p}, \hspace{0.4cm}  
\dot{ {\bf p} }= \nabla U( {\bf p} ).
\label{eq:hamilton}
\end{eqnarray}
where a dot above letters denote the time derivative. 
Without loss of generality, we can assume the center of gravity  
remains at the origin:
\begin{eqnarray}
 \sum_{i=1}^3 m_i {\bf q}_i ={\bf 0}, \hspace{0.4cm} 
\sum_{i=1}^3 {\bf p}_i = {\bf 0}.
\label{eq:CG}
\end{eqnarray}
All these four equations characterize the planar three-body problem.
 
\subsection{Reduction to our variables}

The dimension of the original phase space is twelve.  
The restriction (\ref{eq:CG}) reduces it to eight.
The four variables out of eight are for the configuration 
space, and the remaining four are for the momentum space.
The restriction (\ref{eq:CG}) is equivalent with the fact that 
the two sets of the three vectors form two triangles. 
In the configuration space, there remain two variables. 
Obviously, these are the size and orientation of a triangle. 
Here, the orientation means the direction angle of a selected edge 
with respect to the coordinate axis.  
In the momentum space, there also remain two variables. 
These are the size and orientation of a triangle.  

Let us now look for the dimension of the space 
in which all the independent orbits are contained. 
Here, we say that two orbits are dependent (resp. independent) 
if they can (resp. can not) transform to 
each other under coordinate and time transformations. 
Let us express the four variables in the configuration by 
$({\bf f}, r, \omega)$, where and ${\bf f}$ is a two dimensional vector
and represents the form of the triangle, $r$ is the size, and 
$\omega$ is the orientation. In a similar manner, we express 
the four variables in the momentum space by $({\bf F}, R, \Omega)$. 
Thus we have eight variables $({\bf f}, {\bf F}, r, R, \omega,
\Omega)$. 

Now, consider two sets of variables 
$({\bf f}, {\bf F}, r, R, \omega, \Omega)$ and  
$({\bf f}, {\bf F}, r', R', \omega, \Omega)$. 
If  $r' = \alpha r$ and $R' = \beta R$ for appropriate constants 
$\alpha$ and $\beta$ (this represents a scale transformation), 
motions starting at the two initial conditions are not independent. 
If we choose a particular transformation $r, R \rightarrow 1, R^*$, 
then the variables reduce to 
$({\bf f}, {\bf F}, 1, R^*, \omega, \Omega)$, that is, 
$({\bf f}, {\bf F}, R^*, \omega, \Omega)$. 
 
Two orientations $\omega$ and $\Omega$ are not independent. 
In fact, $\omega$ is measured with respect to a fixed direction in 
the configuration space, and $\Omega$ is measured with respect 
to a fixed direction in the momentum space.
However, the axes in the momentum space can be adjusted to those in 
the configuration space. Let us consider two sets of variables 
$({\bf f}, {\bf F}, R^*, \omega, \Omega)$ and 
$({\bf f}, {\bf F}, R^*, \omega', \Omega')$. 
Then, as is easily understood, if 
$\Omega - \omega = \Omega' - \omega'$ (rotation), motions starting at 
the two initial conditions are not independent. If we choose 
a particular rotation $\omega, \Omega \rightarrow 0, \Omega^*$, 
variables reduce to 
$({\bf f}, {\bf F}, R^*, 0, \Omega^*)$, that is,  
$({\bf f}, {\bf F}, R^*, \Omega^*)$. 
This is our variables.  

In the following two subsections, 
we will carry out the above program to the final set of variables.

\subsection{From Shape Plane to Shape Sphere}

The variables in the position space in this subsection is equivalent 
with those for the shape sphere \cite{Moeckel88} \cite{eight2000} and 
for the FFP \cite{Agekyan}. In \cite{Moeckel88}, the shape sphere has 
been defined as the sphere of constant moment of inertia originally 
introduced by McGehee \cite{McGehee74}.  
Our definition is slightly different from it.  
Our construction of the shape sphere is naive and intuitive.

%The shape sphere is where all the shapes of the triangle 
%formed with a given set of point masses are represented. 
%The space is divided by three meridians of isosceles shape and 
%the equator of collinear shape. 
%The initial condition space of the FFP (for the equal mass case)
%is defined as a bounded area on the plane. It corresponds to 
%a curved triangular domain bounded by the arcs of collinear 
%and isosceles circles. 

We here connect the representations of triangles on the plane and on 
the sphere. 
In order for this, let us introduce the {\it shape plane} as follows. 
Let mass points $m_2$ and $m_3$ be at $B(-0.5, 0.0)$ and $C(0.5, 0.0)$, 
respectively in the $(x,y)$ plane and the position of $m_1$ be $P(x,y)$. 
If $m_1$ changes position in the plane, then triangles exhaust 
all of the shape. We call this the shape plane.

Obviously the points $A(0,\frac{\sqrt{3}}{2})$ and 
$A'(0,-\frac{\sqrt{3}}{2})$ correspond to the equilateral configurations. 
The $x$-axis corresponds to collinear configurations, whereas  
the $y$-axis to the isosceles triangles in which the length of 
two edges $\overline{m_1m_2}$ and $\overline{m_1m_3}$ are the same. 
If $m_1$ is on B, C or its distance from the remaining two 
tends infinity, then the shape corresponds to binary collision 
$E_3$, $E_2$ or $E_1$ 
where $E_i$ means that particle $i$ goes to infinity 
or particles $(j,k)$ collide where $i \neq j$ and $i \ne k$.  
If three masses are equal and $m_1$ is at $D(-3/2, 0)$, 
$O(0, 0)$ or $E(3/2, 0)$, then  the corresponding 
configurations is $C_2$, $C_1$ or $C_3$ where 
$C_i$ represents the isosceles or collinear configuration.

%\begin{figure}[htbp]
% \begin{center}
%\makebox[12cm][l]
%{
%\special{epsfile=shape.eps  hscale=0.3}}%
%\label{shape}
%\vskip10cm
%\end{center}
%\caption{The relation between shape plane and shape sphere}
%\end{figure}

\begin{figure}[htbp]
\includegraphics[width=\linewidth]{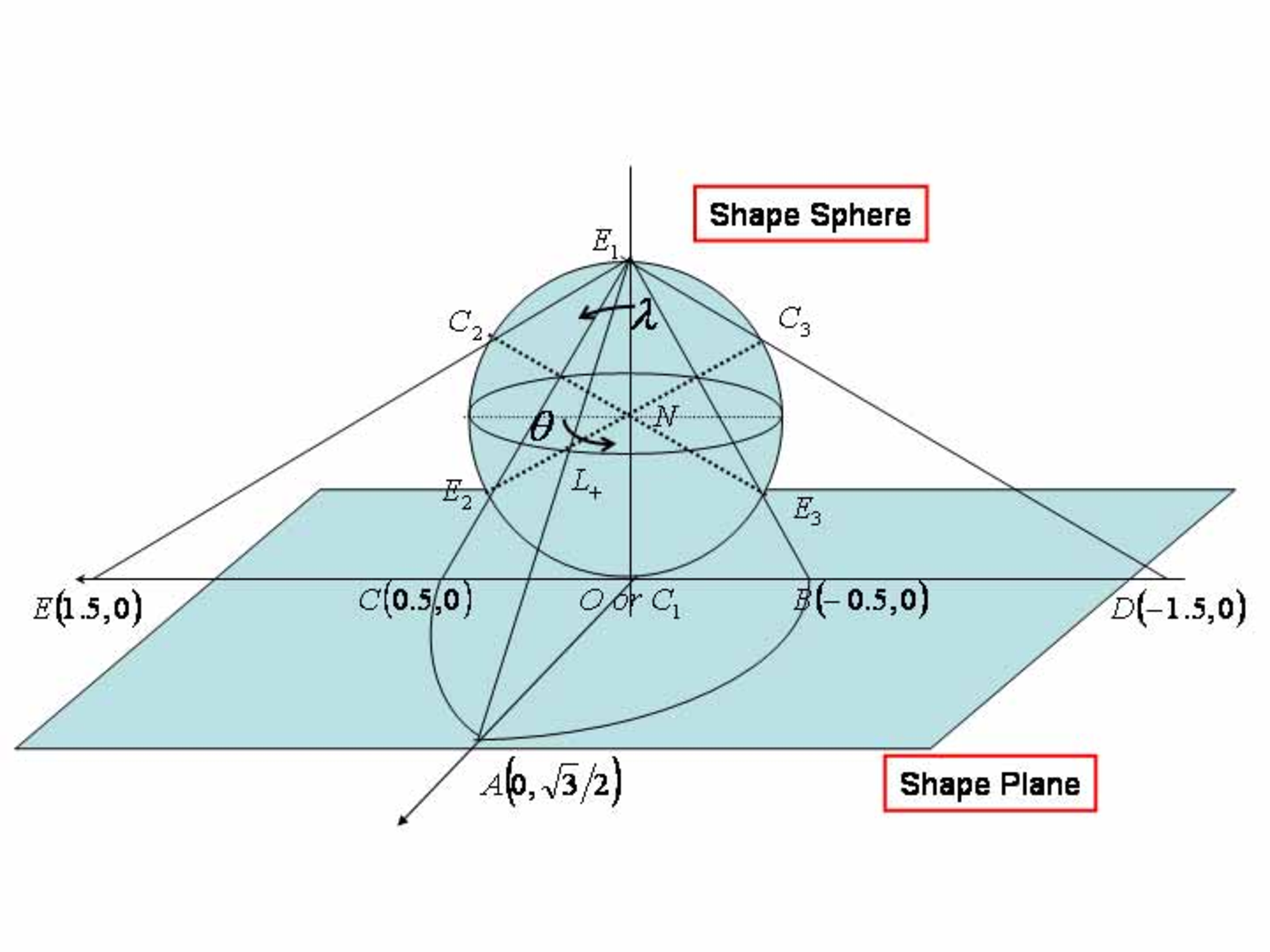}
\caption{The relation between shape plane and shape sphere.}
\label{shape}
\end{figure}

Now, let us obtain the transformation between $(x,y)$ 
on the shape plane and $(\lambda, \theta)$, the longitude and latitude, 
on the shape sphere. We put the sphere of radius $\frac{\sqrt{3}}{4}$ 
in the three-dimensional space $(x,y,z)$ with center at 
$N(0, 0, \frac{\sqrt{3}}{4})$ as in Fig. \ref{shape}. 
The equation for the shape sphere is 
\begin{eqnarray}
 \frac{x^2}{(\sqrt{3}/4)^2} + \frac{y^2}{(\sqrt{3}/4)^2}
+\frac{(z-\sqrt{3}/4)^2}{(\sqrt{3}/4)^2}=1.
\label{eq:sphere}
\end{eqnarray}
Every straight line connecting $E_1( 0, 0, \frac{\sqrt{3}}{2})$ 
and $( x, y, 0)$ meets the sphere at a point. 
Hence every point on the shape plane is mapped to a point on the shape 
sphere, and vice versa. (As usual, infinity in the $(x,y)$ plane is
treated as a point.) 
We denote several particular points on the shape sphere as follows. 
$L_+$ is the intersection of the line $\overline{E_1 A}$ with the sphere. 
$L_-$, $E_3$, $E_2$, $C_2$ and $C_3$, respectively, are the
intersections of 
$\overline{E_1 A'}$, $\overline{E_1 B}$, $\overline{E_1 C}$, 
$\overline{E_1 D}$, $\overline{E_1 E}$ and $\overline{E_1 O}$ with the 
sphere. We denote the center of the sphere by $N$.
We can easily calculate that the angle $\angle E_1 N E_3 $ 
is $2\pi/3$. Other angles can be calculated.  

Let us give this sphere the coordinates $(\lambda, \theta)$, the
longitude and latitude. 
The domains of the definition are $[0, 2\pi]$ and $[0, \pm \pi/2]$.
The origin of $\theta$, the equator, is the great circle in the $xz$-plane.
The north pole is at $L_+$. The origin of $\lambda$ is the great circle
passing through $E_1$ and $L_+$, and $\lambda=0$ at $E_1$. 

Let us take any point $P(x,y,0)$ in the shape plane and connect 
$P$ and $E_1$ with a straight line. Let $P_1(x_1, y_1, z_1)$ be the 
unique intersection of the line and the sphere (6) other than $E_1$.  
Introducing $r = \sqrt{x^2 + y^2}$, we have 
\begin{eqnarray}
 x_1 = \frac{3r}{4r^2+3} \cos \theta,\ 
y_1 = \frac{3r}{4r^2+3} \sin \theta,\nonumber\\ 
\ z_1= \frac{\sqrt{3}}{2} 
\biggl( 1 - \frac{3}{4r^2+3} \biggr), 
\label{X-x}
\end{eqnarray}
and 
\begin{eqnarray}
 \tan \theta =\frac{y}{x}.
\label{theta-x}
\end{eqnarray}
If we represent the $(x_1,y_1,z_1)$ in terms of  
$(\lambda, \theta)$, we have 
\begin{eqnarray}
 x_1 = \frac{ \sqrt{3} }{4} \sin \lambda \cos \theta, \ 
 y_1 = \frac{ \sqrt{3} }{4} \sin \lambda \sin \theta, \nonumber\\
 z_1 = \frac{ \sqrt{3} }{4} ( 1 + \cos \lambda ).
\label{X-lambda}
\end{eqnarray}
From Eqs. (\ref{X-x}) and (\ref{X-lambda}), we get 
\begin{eqnarray}
 \sin \lambda = \frac{ 4\sqrt{3} \sqrt{x^2+y^2} }{ 4 (x^2+y^2) + 3 }.
\label{lambda-x}
\end{eqnarray} 
Thus, Eq. (\ref{theta-x}) and Eq. (\ref{lambda-x})
represent the transformation between $(x,y)$ and $(\lambda, \theta)$. 

In the above, we have moved from the shape plane to the shape sphere 
in the configuration space. Now, in the momentum space, let us move from 
the shape plane to the shape sphere. We need some preparations. 
We know that three momentum vectors make a triangle by
Eq. (\ref{eq:CG}). 
We call this triangle the {\bf momentum triangle}.
We express this triangle in the $(\xi,\eta)$ plane where the $\xi$- and 
$\eta$-axes correspond, respectively, to the $x$- and $y$-axes in Fig. 1. 
We borrow the notation from Fig. 1. 

We normalize the length of ${\bf p}_2$ so that $|{\bf p}_2|=1$. 
Further, we rotate ${\bf p}_2$ so that it aligns with the $\xi$-axis. 
Finally, we put the starting point of vector ${\bf p}_2$ at $B(-0.5,0)$ 
and the end-point at $C(0.5,0)$ (see Fig. 1). We denote this vector by 
$\widetilde{\bf p}_2$. Correspondingly, ${\bf p}_3$ and ${\bf p}_1$ 
are transformed to $\widetilde{\bf p}_3$ and $\widetilde{\bf p}_1$. 
Then, $\widetilde{\bf p}_3$ starts at the end-point of 
$\widetilde{\bf p}_2$, $\widetilde{\bf p}_1$ starts at the end-point of 
$\widetilde{\bf p}_3$, and the end-point of $\widetilde{\bf p}_1$
returns to $B$. 
\begin{eqnarray}
 \widetilde{\bf p}_1 = \overrightarrow{PB},\ 
 \widetilde{\bf p}_2 = \overrightarrow{BC}, \ 
 \widetilde{\bf p}_3 = \overrightarrow{CP}.
\end{eqnarray}

The $(\xi,\eta)$ plane can be called the shape plane of 
momentum triangle. On this plane, we put the sphere of radius 
$\sqrt{3}/4$ as in Fig. 1. We introduce the longitude $\Lambda$ and 
latitude $\Theta$ on this sphere. 
Then the transition from the shape plane $(\xi, \eta)$ 
to the shape sphere $(\Lambda, \Theta)$ in the momentum space can 
be carried out perfectly similar to the case of the configuration 
space.  
We obtain the transformation equations (\ref{theta-x}) and 
(\ref{lambda-x})  with $(\Lambda, \Theta)$ instead of 
$(\lambda, \theta)$ and with $(\xi, \eta)$ instead of $(x,y)$. 

 In the next subsection we give this triangle the size and 
orientation.

\subsection{The remaining variables}
 
In subsection 3.2, we obtained a set of variables 
$({\bf f}, {\bf F}, R^*, \Omega^*)$. In section 3.3, 
we expressed ${\bf f} = (\lambda, \theta)$ and 
${\bf F} = (\Lambda, \Theta)$. In this subsection, we change  
$R^*$ to a more convenient variable. We briefly talk about $\Omega^*$ 
in the last part of this section. 

$R^*$ represents the relative size of configuration and momentum 
triangles. This is obviously related to the relative magnitude of 
the (absolute value of) potential energy $U$ and the kinetic 
energy $T$ (see Eqs. (\ref{eq:energy}) and (\ref{eq:potential})). 
For a given energy and a given configuration of three bodies,  
the configuration triangle is smaller if $U$ is larger, 
whereas, the momentum triangle is smaller if $T$ is smaller.
This consideration makes it plausible to use virial ratio $k$ 
to parametrize the relative size of configuration and momentum 
triangles. $k$ is defined by 
\begin{eqnarray}
 k \equiv  \frac{T}{U} .
\end{eqnarray}

There are two advantages in using the virial ratio as one of the variables.
One is that the global property of the system can be easily grasped. 
In fact, the total energy $h$ of the system is positive if $k>1$, 
whereas $h$ is negative if $k <1$.
The other advantage is that any triple system with negative 
(resp. positive) total energy can be brought to a system with a fixed 
negative (resp. positive) energy by a similarity transformation.
Indeed, we used in \S 3.2 a scale transformation when we normalize 
the size of triangles.  In that transformation, it turns out 
$\beta = \alpha^{-1/2}$. This transformation is equivalent with that of 
the total energy. Let us use this fact to connect $k$ and $R^*$. 

Following the notation of \S 3.2 and \S 3.3, we have 
$r = |\bf{q}_2-\bf{q}_3|$ and $R = |\bf{p}_2|$.  
When we change the scale $r,R \rightarrow 1, R^*$, 
then the relation between these variables is $R^*=\sqrt{r} R$.
The relation between $k$ and $r,R,R^*$ is 
\begin{eqnarray}
 k=\frac{ T(R,\Lambda,\Theta) }{ U(r,\lambda,\theta) }
  =\frac{ T(R^*,\Lambda,\Theta) }{ U(1,\lambda,\theta) },
\end{eqnarray}
 where $T$ and $U$ are represented as functions of 
$(R,\Lambda,\Theta)$ and $(r,\lambda,\theta)$. 

We not yet determine the relative angle of the $x$-axis and
$\xi$-axis. This is related to the starting position of measuring 
angle $\Omega^*$ between configuration and momentum triangles. 
We take the $\xi$-axis (i.e., $\widetilde{\bf p}_2$) 
along ${\bf q}_2$. $\Omega^*$ is then the angle between 
$\widetilde{\bf p}_2$ and ${\bf p}_2$ or the angle between 
${\bf q}_2$ and ${\bf p}_2$, so   
\begin{eqnarray}
 \frac{ \bf{p}_i }{|\bf{p}_i|} = {\cal R}(\Omega^*) 
\frac{ \widetilde{\bf p}_i }{|\widetilde{\bf p}_i|}
={\cal R}(\Omega^*) 
\frac{ {\bf q}_i }{|{\bf q}_i|}
\end{eqnarray}
where ${\cal R} (\Omega)$ is the rotational matrix around the $z$-axis 
by $\Omega$.

Finally, we have  six variables for the planer three-body problem 
$(\lambda, \theta, \Lambda, \Theta, k, \omega)$ where we use $\omega$ 
instead of $\Omega^*$. This space is 
$S^2 \times S^2 \times (I \times S^1)$ where $I = [0, 1)$ or 
$I=[0, \infty]$. If we consider systems with negative energy, then
$I = [0, 1)$ and all variables are bounded.

\subsection{Example orbits in the shape space}

 Let us represent some preiodic orbits in our shape space. 

 Collinear motions are on the equators of the configuration and 
momentum shape spheres, and $\omega$ is equal to $0$ or $\pi$, 
whereas $k$ changes with time. 
We know that the isosceles motions move on the meridians of the 
configuration shape sphere. These motions are represented also 
on the meridians of the momentum shape sphere.  

 The position $(\lambda, \theta)$ of the Euler collinear motion
is fixed on the equator of the configuration shape and changes
depending on the mass ratio of bodies. 
 Its position  $(\Lambda,\Theta)$ is also fixed  
depending on the mass ratio and $(\lambda, \theta)$, 
and generally $\Theta \neq 0$.
The Lagrange motion is fixed at the north or south 
pole of the configuration shape sphere (See th ecross in Fig. 2). 
$(\Lambda,\Theta)$ is fixed and depends only by the mass
ratio. Generally $\Theta \neq 0$. See Fig.3 for the motion of 
$(k,\omega)$. The loci are concentric closed curves with center at 
$k=0.5, \omega=\pi/2$ which is the position of the so-called Lagrange
solution. 

\begin{figure}[htbp]
   \rotatebox[origin=c]{-90}{
\includegraphics[width=0.8\linewidth]{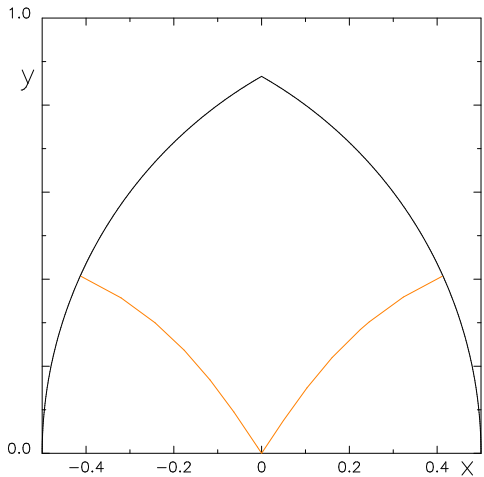}
}
\caption{The figure for the orbit of figure-eight on the xy-plane.}
\label{eight}
\end{figure}

\begin{figure}[htbp]
    \rotatebox[origin=c]{-90}{
\includegraphics[width=0.8\linewidth]{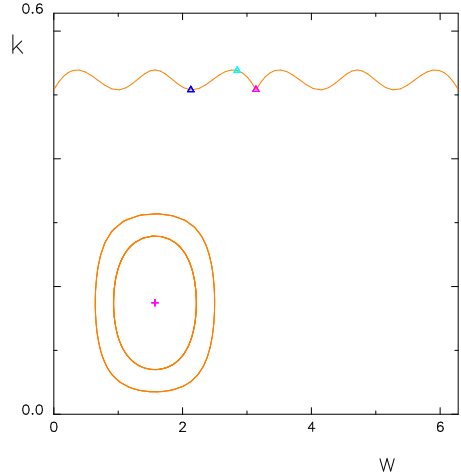}
}
\caption{The figure for the orbit of figure-eight and lagrange with
 k=0.1, 0.2, 0.5 on the kw-plane.}
\label{manykw}
\end{figure}

In the case of the famous figure-eight solution, the motion is  
represnted by the identical curves in the $\lambda$--$\theta$ sphere and 
$\Lambda$--$\Theta$ sphere because of the similarity of configuration 
and momentum triangles \cite{Fuji2004}. We show this orbit 
on the $xy$-plane in Fig. \ref{eight}.
Its $k$-$\omega$ motion is interesting.
We show these in the Fig. \ref{manykw}. Blue and red triangles
correspond to collinear configuration, whereas a green triangle to 
isosceles configuration. 

In the above three examples, it is interesting to note that these orbits 
respectively have the same trajectories in the configuration and momentum 
spheres. The difference manifest itself in the $(k,\omega)$-surface. 
In a sense, these orbits are degenerate. We expect that general periodic 
orbits have different trajectories in three surfaces of the shape
space.

\section{The final step to our variables}

The space represented by variables 
$(\lambda, \theta, \Lambda, \Theta, k, \omega)$
can be interpreted in two different ways. The first interpretation 
is that this is a {\it shape space}. This is an extension  
of the shape sphere (\cite{Moeckel88}, \cite{eight2000}). 
The other interpretation is that this is an initial condition space. 
Any possible initial condition of the planar three-body
problem can be expressed in this space.

\subsection{Global surface of section}

In this section, we try to further decrease the number of variables 
by one. This can be realized if we find a global surface of section. 
To find a global surface of section is motivated 
from the rectilinear problem (\cite{MH1989}). 
We suggest the surface $\dot{I} = 0$, where $I$ is the moment of inertia 
of the triple system and $\dot{I}$ is its time derivative. 

We may change variables from $(\lambda, \theta, \Lambda, \Theta, k, \omega)$ 
to $(\lambda, \theta, L, \dot{I}, k, \omega)$ 
where $L$ is the total angular momentum 
of the system. The equations for the transformation are
\begin{eqnarray}
 L = \sum_i {\bf q}_i \wedge ( {\cal R}(\omega) {\bf p}_i ) 
   =  R^* \sum_i {\bf q}_i \wedge ( {\cal R}(\omega) {\bf p}_i )  \\
 \dot{I} = \sum_i {\bf r}_i \cdot ( {\cal R}(\omega) {\bf p}_i ) 
   =  R^* \sum_i {\bf r}_i \cdot ( {\cal R}(\omega) {\bf p}_i ) ,  
\end{eqnarray}
where ${\cal R}(\omega)$ is the rotation matrix used in \S 3.4.

Now we state the following result. Then we can move to the 
hyper-surface $(\lambda, \theta, L, \dot{I}=0, k, \omega)$. 
This hyper-surface plays a role of 
the global surface of section in the sense that almost all (in the 
sense of measure) orbits pass through this surface. 

\vspace{0.3cm}
\noindent
{\bf Proposition}. Orbits except those of measure zero experience 
$\dot{I}=0$.

\vspace{0.3cm}
\noindent
{\it Proof.} Let us first recall the classification of \cite{Chazy1922}: 

\begin{tabular}{rl}
  H; &hyperbolic motions of three bodies,\\
 HE$_i$; &hyperbolic-elliptic motions \\
         &in which particle $i$ escape, \\
 P; & parabolic motions of three bodies,\\
 PE$_j$; & parabolic-elliptic motions \\ 
	 & in which particle $j$ escape, \\
  B; & bounded motions, \\
  OS; & oscillatory motions. \\
\end{tabular}

\vspace{0.3cm}
These motions can be initial motions and final motions. 
Then there are 36 combinations of initial and final motions such as 
H$^-$ -- B$^+$ where '$-$` indicates the initial motion and '$+$` 
indicates the final motion. 
Among these, combinations 
of escape initial motions and escape final motions give necessarily 
$\dot{I}=0$ because $I \rightarrow \infty$ as 
$t \rightarrow \pm \infty$ and $I$ should attain a minimum at some 
finite $t$. 
In the case of OS, there is a minimum of $I$ in each oscillation, so  
there is a time such that $\dot{I}=0$.

The remaining combinations to be examined are between $B$ and escape 
motions and between $B$ itself. However, the combinations of $B$ and 
escape motions occupy a set of measure zero in the phase space, as 
has been shown (\cite{V.M.Alekseev}). 
Therefore we only need to check the last combination, B$^-$ -- B$^{+}$. 

If the orbit does not experience $\dot{I}=0$ 
in the range $t_0 < t < \infty$, 
then it means $\dot{I} >0$ or $\dot{I}<0$ in this range of $t$. 
We put $\dot{I}_{\sup}=\sup_{t_0 < t < \infty}{\dot{I}}$ and 
$\dot{I}_{\inf}=\inf_{t_0 < t < \infty}{\dot{I}}$.
Then for the case $\dot{I}>0$, 
\begin{eqnarray}
 I(t) &=& \int^t_{t_0} \dot{I}(t) dt 
> I(t_0) +  \dot{I}_{\inf} (t - t_0),  \label{key:eq22}
\end{eqnarray}
and for the case $\dot{I}<0$, 
\begin{eqnarray}
 I(t) &=& \int^t_{t_0} \dot{I}(t) dt 
< I(t_0) - |\dot{I}_{\sup}| (t-t_0) \label{key:eq23}
\end{eqnarray}

Let us first consider the case (\ref{key:eq22}). In order that 
$I(t)$ remains finite as $t \rightarrow \infty$, it is necessary 
that $\dot{I}_{\inf}=0$. This means that the triple system 
asymptotically tends to 
a configuration with constant $I$ as $t \rightarrow \infty$. 
In order that $I(t)$ remains 
greater than zero in the case (\ref{key:eq23}) as $t
\rightarrow  \infty$, it is necessary that $\dot{I}_{\sup}=0$.
In this case, the triple system asymptotically tends to a configuration 
with constant $I$ as $t \rightarrow \infty$. 
If $I(t) \rightarrow 0$ as $t$ increases, 
this corresponds to triple collision with a finite collision time
$t^*$. In this case, the system does not experience $\dot{I} = 0$ for
$t_0 < t < t^*$. As is well known, orbits which experience triple 
collision occupy a set of zero-measure in the phase space.

The remaining problem is to estimate the measure of the orbits 
which asymptotically tend to $\dot{I} =0$ as $t \to \infty$. 
These orbits constitute the stable set of 
orbits on the set $\dot{I} =0$. In any situation, this stable set 
does not span the phase space since otherwise contradiction to the volume
preservation will be derived. Therefore, the set of orbits which does
not experience $\dot{I} =0$ is of measure zero. 
\hfill Q.E.D.

\subsection{An extension to the three-dimensional case}

Before extending our results to the three dimensions, 
we discuss on the dimension of the initial value space of the
three-dimensional three-body problem.
There are ten equivalence relations.
Six of these are for the translations,  
three of these are for the rotation, 
and one of these are for the scale.
Each of them has its equivalence class. 
We denote these sets by ${\bf E}_i (i=1,2,...,10)$, 
${\rm dim}~{\bf E}_i=1$ 
and ${\bf E}_i \cap {\bf E}_j  = \emptyset$. 
We define the set ${\bf X}$ as ${\bf R}^{18}/(\Pi_{i=1}^{10} {\bf E}_i)$,
then the dimension of ${\bf X}$ is 8. 
We will find eight bounded variables in ${\bf X}$ in what follows. 

Even in the three-dimensional case, 
we can define the plane in which the three bodies live.
We take this plane as the initial configuration plane. 
However, in contrast to the planar case, the plane in which 
the configuration triangle exists and the plane in which 
the momentum triangle exists moves separately.
Thus we need to specify the relative position of two planes.

%\begin{figure}[htbp]
%\includegraphics[width=\linewidth]{threeD.eps}
%\caption{The image for the three angle variables.}
%\label{threeD}
%\end{figure}

Now, we look for the remaining variables. 
Let us consider an arbitrary state of the triple system, 
and suppose that the shapes and the sizes of configuration and
momentum triangles are determined.  
As before, we put the center of gravity of the configuration 
triangle at the origin of the $(x,y)$-plane, and 
put ${\bf q}_3 - {\bf q_2}$ along the $x$-axis.   
We here define the nominal position of the momentum triangle. 
Let $\widetilde{\bf p}_i$ be the momentum of particle $i$ in 
the nominal position. We put the center of mass of the triangle made 
with $\widetilde{\bf p}_i$ at the origin of the $(x,y)$-plane, and  
put $\widetilde{\bf p}_2$ parallel to ${\bf q}_2$. We take 
the $\xi$-axis in this direction, and take the $\eta$-axis perpendicular
to the $\xi$-axis. The $\zeta$-axis is defined as the third coordinate 
axis in the momentum space. 
We denote the $(\xi,\eta)$-plane by $\pi_0$. 
So, initally, the plane of the nominal momentum triangle coincides with 
the plane of the configuration triangle. 
 
In order to bring the nominal momentum triangle to the the actual 
momentum triangle, we need three steps. 

\begin{enumerate}
 \item[(1)] Rotate $\pi_0$ around the $\zeta$-axis by angle $\omega$. 
We call the new plane $\pi'$, 
and denote the $\xi$-axis of $\pi'$ by $\xi_{\pi'}$.
 \item[(2)]  Rotate $\pi'$ around the $\xi_{\pi'}$
by angle $\psi$. We call the new plane $\pi''$. Denote the $\zeta$-axis 
of $\pi''$ by  $\zeta_{\pi''}$. 
 \item[(3)]  Rotate $\pi''$ around $\zeta_{\pi''}$ by angle $\phi$. 
Then the triangle coincides with the actual momentum triangle. 
\end{enumerate}

\noindent
Here, the angle $\omega$ coincides with that in the two-dimensional 
case.

Now it is easy to make the initial conditions 
for the three-dimensional case.
At first, we make the two-dimensional initial condition for the form 
of the {\it  momentum triangle} at the nominal place.  
Second, we rotate the plane to the position where the actual 
momentum triangle exists.
Then the eight variables for the three-dimensional case are  
$(\lambda ,\theta, \Lambda, \Theta, k, \omega, \psi, \phi)$. 
As in the two dimensional case we can transform variables 
from $( \lambda, \theta, \Lambda, \Theta, k, \omega, \psi, \phi)$ to 
$( \lambda, \theta, L, \dot{I}, k, \omega,\psi,\phi)$,
where $L$ and $\dot{I}$ are 
the total angular momentum and the derivative of the total 
moment of inertia. Finally, the surface defined by the $\dot{I}=0$ 
will be the global surface of section as in the two-dimensional case.

\section{Summary and Discussion}

In this report, for a given mass, we have extended the free-fall problem 
(FFP) to the full three-body problem of three dimensions.
We find new variables $(\lambda, \theta, L, \dot{I}, k, \omega,\phi,\psi)$ 
for the three-dimensional case and 
$(\lambda, \theta, L, \dot{I}, k, \omega)$ 
for the two-dimensional case which are 
convenient for computer simulations.
The reasons for it are the following. 

\begin{list}{}{}
 \item[(1)] If the virial ratio $k$ is positive and large, 
then the total energy of the triple system is positive and its final 
motion is simple. So we can omit this case from our consideration, and 
then we can bound the value of $k$. 
 \item[(2)] If we set the virial ratio $k \le 1$, then the domain of 
the definition for $L$ and $\dot{I}$ is bounded. 
 \item[(3)] The surface defined by $\dot{I}=0$ in the space 
$(\lambda,\theta,L,\dot{I},k,\omega,\phi,\psi)$ becomes 
the global surface of section. 
We can map the structure of the whole phase space on this surface of section.
%\item[(4)] It is obviously that the ranges of the angle variables is bounded.
\end{list}

Let us briefly discuss about the future role of the present result. 
The progress of computer enables us to calculate immense number of orbits. 
We can see the projections of the phase space through integrations of
orbits by fixing some of the variables such as $k$, $L$, and $\dot{I}$ 
in the initial value space. 

As seen in \S 3.5, known periodic orbits occupy special positions in 
the shape space. We hope that new kinds of periodic orbits may be found 
with the aid of this shape-space representation. 

%Finally, the three-tangents theorem \cite{Kuwabara2006} connects 
%the areas of configuration and momentum triangles and $\omega$
%with the square of the angular momentum 
% (the area of the triangle made by three-tangents is related to 
%the area of the configuration triangle and $\omega$). 

\vspace{0.5cm}
\noindent
{\bf Acknowledgments.} 

One of the authors (K.K.) expresses his thanks to Prof. Keiichi Maeda 
for encouragements. 


{\small 

\begin{thebibliography}{99}


\bibitem{Agekyan} Agekyan, T.A. and Anosova, J.P.: 1968,
{\it Soviet Physics-Astronomy} {\bf 11}, 1006 - 1014.

\bibitem{V.M.Alekseev} Alekseev, V.M.: 1981
{\it Amer.Math.Soc.Transl. } (2) Vol.116.

\bibitem{Anosova1981} Anosova, J.P., Bertov, D.I. and Orlov, V.V.: 1981,
Astrophysics {\bf 20}, 177. 

\bibitem{Anosova1986} Anosova, J.P.: 1986,
{\it Astrophys. Sp. Sci.} {\bf 124} 217 - 241.


\bibitem{Brou} Broucke, R.: 1979, {\it Astron. Astrophys}. {\bf 73}, 
303 - 313.

\bibitem{Chazy1922} Chazy, M.: 1922, {\it Ann. Sci. \'Ecole Norm. Sup.} 
(3) {\bf 39}, 29 - 130.

\bibitem{eight2000} Chenciner, A. and Montgomery, R.: 2000, A remarkable 
periodic solution of the three-body problem in the case of equal masses, 
{\it Ann. Math.} {\bf 152}, 881 - 901. 

\bibitem{HM1993} Hietarinta, J. and Mikkola, S.: 1993,
{\it Chaos } {\bf 3} (2), 1993.

%\bibitem{Kuwabara2005} Kuwabara, K. and Tanikawa, K.: 2005,
%in the Proceedings of Workshop on Few-Body Problem: Theory and 
%Computer Simulations, Turku, Finland, July 3 - 8, 2005.

%\bibitem{Kuwabara2006} Kuwabara, K. and Tanikawa, K.: 2006,
%{\it Phys. Lett. A} {\bf 354}, 445 - 448. 

\bibitem{McGehee74} McGehee, R.: 1974, Triple collision in the 
collinear three-body problem, {\it Invent. Math.} {\bf 27}, 
191 - 227.

\bibitem{MH1989} Mikkola, S. and Hietarinta, J.: 1989,
{\it Celestial Mechanics and Dynamical Astronomy} 
{\bf 46}, 1 - 18.


\bibitem{MS1990} Mikkola, S. and S.V.Aarseth.: 1990,
{\it Celestial Mechanics and Dynamical Astronomy} 
{\bf 47}, 375-390.

\bibitem{MS1993} Mikkola, S. and Aarseth, S.V.: 1993,
{\it Celestial Mechanics and Dynamical Astronomy}
{\bf 57}, 439-459.

\bibitem{Moeckel88} Moeckel, R.: 1988, Some qualitative features of the 
Three-body problem, {\it  Contemporary Mathematics} 
{\bf 81}, 1 - 17.

\bibitem{Tanikawa1995} Tanikawa, K., Umehara, H., and Abe, H.: 1995,
{\it Cel. Mech. Dynam. Astron.} {\bf 62}, 335 - 362.

\bibitem{TM2000a} Tanikawa, K.and Mikkola, S.: 2000,
{\it Cel. Mech. Dynam. Astron.} {\bf 76}, 23 - 34.

\bibitem{TM2000b} Tanikawa, K. and Mikkola, S.: 2000,
{\it Chaos} {\bf 10}, 649 - 657.

\bibitem{Umehara00} Umehara, H. and Tanikawa,K.: 2000,
{\it Cel. Mech. Dynam. Astron.} {\bf 76}, 187 - 214.


\bibitem{Whittaker1952} Whittaker, E.T.: 1952,
{\it A Treatise on the Analytical Dynamics of Particles and 
Rigid Bodies}, Cambridge University Press, Fourth Edition. 

\bibitem{ZaCh} Zare, K. and Chesley, S.: 1998, {\it Chaos } {\bf 8},
475-494.

\bibitem{Fuji2004}T. Fujiwara, H. Fukuda, A. Kameyama,
 H. Ozaki and M. Yamada, 
{\it Journal of Physics A} {\bf 37} (2004), p.10571. 
\end{thebibliography}

}
\end{document}